\documentclass{desyproc}

\begin{document}
	\title{New constraints on very light pseudoscalars}

	\author{{\slshape A.~Payez$^1$, J.R.~Cudell$^{1}$, D.~Hutsem\'ekers$^{1}$}\\[1ex]
	$^1$IFPA group, AGO Dept., University of Li\`ege, Li\`ege, Belgium}


	\desyproc{DESY-PROC-2011-04}
	\acronym{Patras 2011} 
	\doi  

	\maketitle

	\begin{abstract}
		Nearly massless axion-like particles are of interest for astrophysical observations, and some constraints on their parameter space do exist in the literature. Here, we propose to put new constraints on these particles using polarisation and, in particular, the polarisation differences observed between different quasar classes.
	\end{abstract}

		There is quite a number of puzzling phenomena in astrophysics that could be clarified with the existence of very light scalar or pseudoscalar particles, commonly refered to as {\em axion-like particles} (a.k.a. ALPs).
		One of the main reasons is that these hypothetical particles can have a coupling to photons inside external magnetic fields, as in the axion case~\cite{Raffelt:1987im}, and can lead to observable signatures in every measurable property of light: its intensity, its spectrum, or its polarisation.
		In this work, we want to present polarimetry as a tool to search for (or, at least, to constrain) the existence of such nearly massless particles.

		To illustrate this, we use the sample of 355 high-quality polarisation measurements which confirmed the existence of large-scale correlations of (linear) polarisation position angles of visible light from quasars in some regions of the sky~\cite{Hutsemekers:2005iz}. It was thought that these large-scale alignments could be naturally explained by ALP-photon mixing, as their phenomenology predicts the creation of linear ($p_{\mathrm{lin}}$) and circular ($p_{\mathrm{circ}}$) polarisations. However, the recent observation of vanishing circular polarisation~\cite{Hutsemekers:2010fw} is a problem for this hypothesis; see~\cite{Payez:2011sh} and references therein.

		As ALPs cannot explain the alignments, these data can be exploited to constraint axion-like-particle parameters. We are going to consider objects located behind the Virgo supercluster, as we have information on $p_{\mathrm{lin}}$ and $p_{\mathrm{circ}}$ in this region.
One can then consider ALP-photon mixing in the magnetic field of that supercluster, and check for which parameters the generated polarisation would match (or in this case contradict) the observations. We use the following:
		\begin{itemize}
			\item For the type of quasars we are discussing, which have linear polarisations at the 1\%-level, it is known~\cite{Hutsemekers:1998pp} that objects with broad absorption lines (BAL quasars) are typically more polarised than other radio-quiet quasars in the continuum: $\langle p_{\mathrm{lin}}^{\textrm{BAL}} - p_{\mathrm{lin}}^{\textrm{non-BAL}}\rangle \lesssim 2\%$. This feature still holds in the sample considered; the expected distributions of polarisation are thus preserved~\cite{Hutsemekers:2001}. Therefore, while we do not have access to the initial polarisations, we know that the linear polarisation generated by ALP-photon mixing on the way towards us cannot be larger than 2\%, otherwise we would not see the difference between the two quasar distributions anymore.
			\item Even though we know that quasars are intrinsically linearly polarised, we do not want to make assumptions on the initial distribution of polarisations. Therefore, to avoid any overestimation of the final (observed) polarisation, we start from unpolarised light beams. ALP phenomenology tells us that we are going to underestimate the linear, but also the circular polarisation. This, in turn, will give us conservative constraints.
			\item As reported in~\cite{Hutsemekers:2010fw}, out of the 21 $p_{\mathrm{circ}}$ measurements of quasars, all have null circular polarisation, except two highly linearly polarised blazars which might be intrinsically circularly polarised. As we want to be conservative, we use a $3\sigma$ level bound for $p_{\mathrm{circ}}$ and require that the circular polarisation produced by ALP-photon mixing is within these bounds. Also, as we only have data for objects with $p_{\mathrm{lin}}\geq 0.6\%$, we will only apply the $p_{\mathrm{circ}}$ constraint if the linear polarisation is such.
		\end{itemize}

\begin{figure}[h!!]
			\begin{center}
				\includegraphics[width=0.4\textwidth]{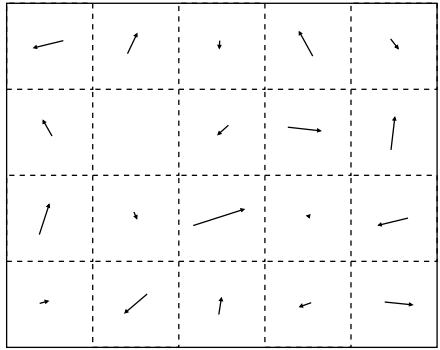}
			\end{center}
			\caption{Sketch of the assumed cellular morphology for the transverse magnetic field in superclusters.}
			\label{fig:patchy}
		\end{figure}

		For cluster and supercluster magnetic fields~\cite{Vallee:2011}, the morphology discussed in the literature is often a ``patchy'' one. More precisely, one can assume a cell-like composition, such that the magnetic field is constant in each cell\footnote{Here, for the sake of simplicity, we will not consider fluctuations; this is done in~\cite{Payez:prep}.} but its direction changes from cell to cell. Here, we allow for 3-dimensional rotations~\cite{Payez:2011sh}, so that when the field picks up a longitudinal component (most of the time), this leads to (much) lower values of the mixing, as only the transverse part of the magnetic field is relevant for ALPs. In the end, taking a slice of the magnetic field projected in the transverse plane, we get something similar to what is depicted in Fig~\ref{fig:patchy}.

		With such a configuration, when looking at objects located in different directions, we effectively look at different fields. As the exact structure of the magnetic field inside the Virgo supercluster is poorly known, we assume that the fields are sufficiently irregular so that, for photons coming from different objects, the encountered fields essentially correspond to random ones (with possibly some additional correlations at the supercluster scale).
\\

		Our constraints are obtained iteratively for each point in the ALP parameters space (mass $m$, and coupling to photons $g$). Each time, we first generate a random configuration for the magnetic field and, for an initially unpolarised light beam, calculate the linear and circular polarisations after propagation. We then change the magnetic field configuration and repeat, always comparing the results to our conservative constraints. With the number of successes over the number of trials, we eventually obtain a probability for the couple $(m,g)$. This probability effectively measures how likely it is that the mass and coupling of the ALP do not contradict the observations if we do not assume anything special about the magnetic field.
		Note that in the following ``excluded at 2$\sigma$'' means that the probability is $5\%$ or lower (probability of $95\%$ not to be compatible with the data), and similarly for 1$\sigma$ and 3$\sigma$.

		\begin{figure}
			\begin{center}
				\includegraphics[width=\textwidth]{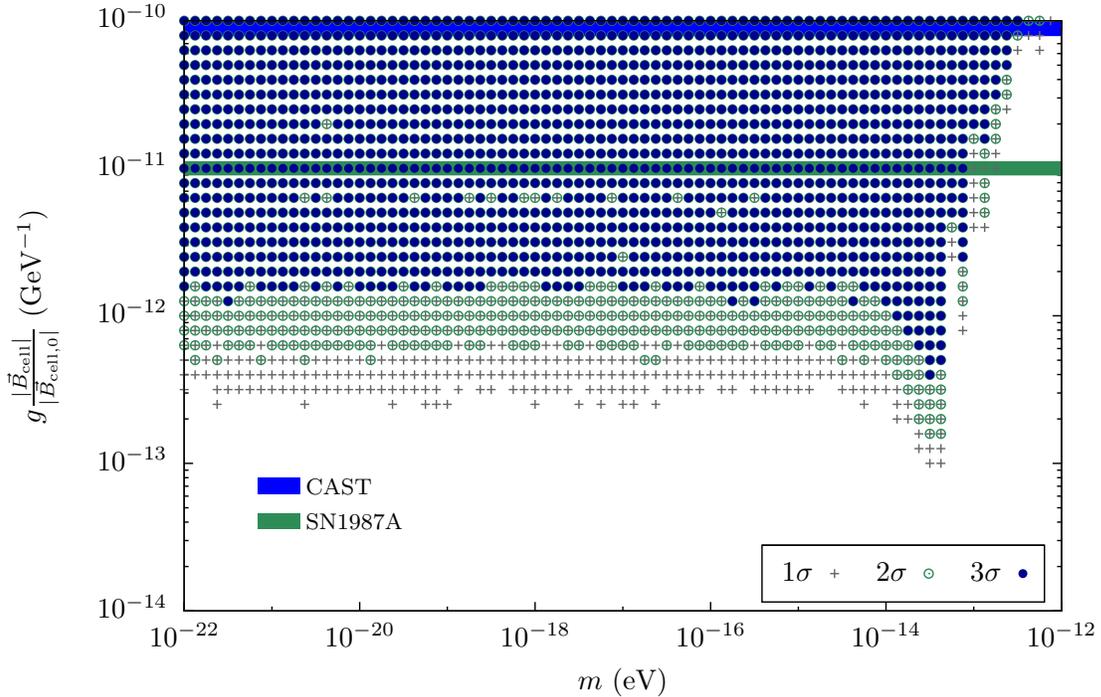}
			\end{center}
			\caption{Exclusion plot for the monochromatic case ($\lambda = 250$~nm) ; parameters are excluded at 1, 2 or 3$\sigma$ CL. Here, $|\vec{B}_{\textrm{cell},0}| = 2~\mu$G and the electron density is $n_e = 10^{-6}$~cm$^{-3}$. Also shown are the constraints, by CAST ($g\lesssim8.8~10^{-11}$~GeV$^{-1}$), and derived from SN 1987A ($g\lesssim10^{-11}$~GeV$^{-1}$), for low-mass ALPs.}
			\label{fig:limits_monochr}
		\end{figure}

		Now, strictly speaking, the observations have been made in a given bandwidth, and the constraints must be applied taking this into account.\footnote{This is presented elsewhere, together with a refined test, leading to even better constraints~\cite{Payez:prep}.} Indeed, as explained in~\cite{Payez:2011sh}, one usually expects a smaller amount of polarisation from axion-photon mixing compared to the simpler monochromatic case. Here, we stick to the mean value of the wavelengths observed (namely, 250~nm)\footnote{Note that, over the supercluster scale, the effect of cosmological redshift on the wavelength is negligible: even though $\lambda$ was shorter at the emission, it does not change over the last 10~Mpc we are interested in.} because it illustrates what constraints we would have if the same observations were performed in a very narrow band with the same statistics.
		If one uses the same method, the constraints are slightly more stringent in this case, as more polarisation is produced.

		The result is seen in Fig.~\ref{fig:limits_monochr}: for low-mass axion-like particles, we can obtain new constraints at the 2$\sigma$-level down to values below $g\approx10^{-12}$~GeV$^{-1}$. For reference, the current best limits, by CAST~\cite{Andriamonje:2007ew}, and derived from SN 1987A~\cite{Raffelt:2006cw}, are shown and are confirmed for nearly massless ALPs.
		For the magnetic field~\cite{Vallee:2011}, we have used a hundred 100~kpc cells with a field strength of 2~$\mu$G; one can also find slightly lower values in the literature for superclusters (down to \mbox{$\approx0.3~\mu$G}), but then typically over larger coherence lengths (up to the supercluster size). The exact value of the magnetic field strength one uses is not really an issue however, as the equations for the mixing always involve the coupling and the magnetic field strength together: as in this case there is only one scale (set by the modulus of the total magnetic field of each cell $|\vec{B}_{\textrm{cell}}|$), Fig.~\ref{fig:limits_monochr} can be rescaled straightforwardly for any other value of $|\vec{B}_{\textrm{cell}}|$ if our best knowledge of it changes.
		Note that the region near $m=10^{-14}$~eV is more significantly excluded as the mixing is stronger for masses close to the plasma frequency.

		We have verified the stability of our constraints against changes of a factor two up and down for the sizes and numbers of cells, or for the electron density, against the effect of an additional background field of $0.3~\mu$G (in which case, one cannot rescale anymore), the influence of fluctuations of the magnetic field strength and of the electron density, and of the average over the observed bandwidth.

	To conclude, we have illustrated that polarimetry is a tool that can be used to probe the ALP parameter space. Using quasar polarisation data and sticking to a conservative approach to avoid any overestimation of the polarisation, we have shown that new constraints can be put on very light axion-like particles. While the constraints are proven to be robust, the only drawback is associated with the uncertainties about the magnetic field of the supercluster. This technique could be applied to observations of polarisation for objects located behind clusters of galaxies, for instance, for which we have a much better knowledge of the magnetic field structure, and to higher-energy data, for instance in X-Ray polarimetry with tools like the GEMS satellite, which is to be launched in 2014 and which will definitely help the search for signals of low-mass ALPs.


	\begin{footnotesize}

	\end{footnotesize}



\begin{thebibliography}{99}

			\bibitem{Raffelt:1987im}
			  G.~Raffelt \& L.~Stodolsky,
			  ``Mixing of the Photon with Low Mass Particles,''
			  Phys.\ Rev.\  D {\bf 37} (1988) 1237.

			\bibitem{Hutsemekers:2005iz}
			  D.~Hutsem\'ekers, R.~Cabanac, H.~Lamy and D.~Sluse,
			  ``Mapping extreme-scale alignments of quasar polarization vectors,''
			  Astron.\ Astrophys.\  {\bf 441} (2005) 915
			  [arXiv:0507274 [astro-ph]].

			\bibitem{Hutsemekers:2010fw}
			  D.~Hutsem\'ekers, B.~Borguet, D.~Sluse, R.~Cabanac and H.~Lamy,
			  ``Optical circular polarization in quasars,''
			  Astron.\ Astrophys.\ {\bf 520} (2010) L7 
			  [arXiv:1009.4049 [astro-ph]].

			\bibitem{Payez:2011sh}
			  A.~Payez, J.R.~Cudell and D.~Hutsem\'ekers,
			  ``Can axionlike particles explain the alignments of the polarizations of
			  light from quasars?,''
			  Phys.\ Rev.\ D {\bf 84} (2011) 085029
			  [arXiv:1107.2013 [astro-ph]].

			\bibitem{Hutsemekers:1998pp}
			  D.~Hutsem\'ekers, H.~Lamy and M.~Remy,
			  ``Polarization properties of a sample of broad absorption line and gravitationally lensed quasars,''
			  Astron.\ Astrophys.\ {\bf 340} (1998) 371;
			  G.D.~Schmidt and D.C.~Hines,
			  ``The Polarization of Broad Absorption Line QSOs,''
			  Astron.\ J.\ {\bf 512} (1999) 125;
			  H.~Lamy and D.~Hutsem\'ekers,
			  ``Polarization properties of broad absorption line QSOs: New statistical
			  clues,''
			  Astron.\ Astrophys.\  {\bf 427} (2004) 107
			  [arXiv:0408476 [astro-ph]].

			\bibitem{Hutsemekers:2001}
			  D.~Hutsem\'ekers \& H.~Lamy,
			  ``Confirmation of the existence of coherent orientations of quasar
			  polarization vectors on cosmological scales,''
			  Astron.\ Astrophys.\ {\bf 367} (2001) 381
			  [arXiv:0012182 [astro-ph]].

			\bibitem{Vallee:2011}
			See \textit{e.g.} J.P.~Vall\'ee,
			``Magnetic fields in the galactic Universe, as observed in supershells, galaxies, intergalactic and cosmic realms,''
			New Astron.\ Rev.\ {\bf 55} (2011) 91;
			Y.~Xu {\it et al.},
			``A Faraday Rotation Search for Magnetic Fields in Large Scale Structure,''
			Astrophys.\ J.\  {\bf 637} (2006) 19.

			\bibitem{Payez:prep}
			A.~Payez, J.R.~Cudell and D.~Hutsem\'ekers, submitted
			[arXiv:1204.6187 [astro-ph]].

			\bibitem{Andriamonje:2007ew}
			  S.~Andriamonje {\it et al.}  [CAST Collaboration],
			  ``An improved limit on the axion-photon coupling from the CAST experiment,''
			  JCAP {\bf 0704} (2007) 010
			  [arXiv:0702006 [hep-ex]].

			\bibitem{Raffelt:2006cw}
			  See G.~G.~Raffelt,
			  ``Astrophysical axion bounds,''
			  Lect.\ Notes Phys.\  {\bf 741} (2008) 51
			  [arXiv:0611350 [hep-ph]] and references therein.

		\end{thebibliography}
\end{document}